\documentclass[12pt]{iopart}
\usepackage{iopams}  
\usepackage{bm}
\usepackage{graphicx}
\begin{document}
% Journal identifier can be put here if required, e.g.
\SUST

\title[Critical currents in a periodic coplanar
array of narrow superconducting strips]{Magnetic-field dependence of the critical
currents in a periodic coplanar array of narrow superconducting strips}

\author{John R Clem,$^1$\footnote[3]{To whom correspondence should be
addressed.} Ali A Babaei Brojeny$^2$  and Yasunori Mawatari$^3$}

\address{$^1$\ Ames Laboratory and Department of Physics and Astronomy,\\
  Iowa State University, Ames, Iowa, 50011--3160, USA}

\address{$^2$\ Department of Physics, Isfahan University of Technology,
Isfahan 84154, Iran}

\address{$^3$\ National Institute of Advanced Industrial Science and Technology
(AIST)\\
	Tsukuba, Ibaraki 305--8568, Japan}

 \ead{clem@ameslab.gov}

\begin{abstract}We calculate the magnetic-field dependence of the critical
current due to both geometrical edge barriers and bulk pinning in a
periodic coplanar array of narrow superconducting strips.  We find that in zero or
low applied magnetic fields the critical current can be considerably enhanced by
the edge barriers,  but  in modest applied magnetic fields
the critical current reduces to  that due to bulk pinning alone.
\end{abstract}


\submitted{\SUST}

\maketitle

\section{Introduction}

The critical current at which a voltage appears along the length of a
superconductor is one of the most important factors considered
in applications of superconductivity \cite{Campbell72}.  The critical current is
known to depend upon both local pinning centers in the
material and the shape of the  conductor's cross section.  Even
in the absence of bulk pinning, isolated type-II superconducting thin-film
strips subjected to perpendicular magnetic fields show magnetic hysteresis
due to  geometrical edge
barriers \cite{Indenbom94,Schuster94,Zeldov94a,Benkraouda96,Doyle97,
Maksimov98}.  Such strips  have finite critical currents arising
from the edge barriers \cite{Benkraouda98}. It also has been shown that 
the critical currents become larger when slits are fabricated near the
edges of the strips \cite{Mawatari01}; the edge-barrier effects are
enhanced because the slits increase the number of edges that can prevent
flux penetration into the inner strips.  Although at subcritical applied
fields and currents the magnetic response of two parallel strips in the
Meissner state is reversible \cite{Brojeny02}, when the  strips are
connected at their ends and an applied magnetic field exceeds a certain
value, magnetic flux penetrates irreversibly and the magnetic response
becomes hysteretic as a consequence of the edge
barriers \cite{Zhelezina02,Ainbinder03,Maksimova04}. A detailed analysis of the
effects of edge barriers upon the magnetization hysteresis in samples consisting
of one, two, and three parallel strips connected at their ends has been presented
in \cite{Mawatari03}.

To calculate the combined effects of geometrical
barriers and bulk pinning is more difficult, but this has been accomplished for
single strips in
\cite{Zeldov94a,Benkraouda96,Maksimov98,Maksimov95b,Maksimov95a,Elistratov00,
Maksimova01,
Kupriyanov74}, and a theoretical analysis of the magnetic-field dependence of the
critical current of an isolated superconducting strip due to both an edge barrier
and uniform bulk pinning  has been presented in  \cite{Elistratov02}.
Here we extend the above calculations to account for both
geometrical edge barriers and bulk pinning in an infinite number of strips
using the
$X$-array method \cite{Mawatari96}, by which the magnetic-field and
current-density distributions for an array of parallel superconducting
strips arranged periodically along the $x$ axis in the
plane $y=0$ can be obtained analytically from the solutions for an isolated
superconducting strip \cite{Elistratov02}.
In particular, we consider the case for which each strip in the array carries an
equal amount of current in the presence of a perpendicular applied magnetic field.
We then calculate the critical current
$I_c$ of each strip accounting for both a geometrical edge barrier, which impedes
the entry of vortices into the strip, and uniform bulk pinning, which impedes the
motion of vortices across the strip.
 
Our calculation is relevant to a number of recent studies of the ac properties
of striated coated-conductor tapes, i.e., tapes that have been subdivided
into parallel thin strips (filaments) separated by narrow gaps. Ideally, to
minimize ac losses in power engineering applications using multifilamentary
conductors, the individual filaments should not only have small cross sections but
also be twisted and transposed, such that the filaments are inductively
equivalent, are decoupled from each other, and thus carry equal
current \cite{Wilson83}.  In practice, it is not possible to satisfy all these
conditions, and compromises are generally accepted for practical reasons.  Carr
and Oberly \cite{Carr99} have suggested that that the ac
losses in tapes several millimeters wide could be reduced by first subdividing the
tapes by striations and then twisting the tapes. 
Analytical and experimental results pursuing this idea have been reported
in 
\cite{Oberly01,Cobb02,Polak02,
Amemiya04,Barnes04,Levin05,Wang05,Sumption05,Majoros05,Tsukamoto05,Levin06a,
Demencik07,Polak06,Levin06b}.
Ashworth and Grilli \cite{Ashworth06} have recently proposed a variation of this
approach, in which the tape is also  subdivided into narrow parallel filaments, but
instead of twisting the tape, the filaments are interrupted periodically by
transverse cross-cuts bridged with normal metal.  Magnetic flux can enter at the
cross-cuts, thereby decoupling the filaments and allowing equal
currents to flow in each filament.   The additional 
ohmic losses in
the normal bridges are more than compensated by a huge reduction in the far more
important ac coupling losses.

In the present paper we examine the dc properties of an infinite array of parallel
superconducting strips, which can be regarded as an approximation to a striated
coated-conductor tape of finite width.  Our paper focuses on the possibility that a
geometrical edge barrier can increase the critical current of each strip above
that due to bulk pinning alone, but it also should be possible to use our results
in calculating the hysteretic ac losses of an array of parallel strips as in
 \cite{Mawatari96,Mawatari97,Muller97,Muller99}. 

Our paper is organized
as follows.  In Sec.\
\ref{X_arrays} we review our complex-field approach and the
$X$-array method. In Sec.\ \ref{bulk-pinning} we apply the complex-field
approach and the $X$-array method to obtain the complex field and
the critical currents in a periodic array of parallel superconducting
strips subject only to bulk pinning.   In Sec.\
\ref{edge_barrier_only} we calculate the magnetic-field dependence of the
critical current of one of the strips when the infinite array is subject
only to geometrical edge barriers. In Sec.\ \ref{both_effects} we calculate
the critical current of one of the strips when both geometrical edge
barriers and bulk pinning are present.  We discuss the results and present
our conclusions in Sec.\
\ref{conclusions}.

\section{\label{X_arrays}% 
Complex fields and $X$ arrays}

We begin by reviewing the properties of a long superconducting strip
of thickness $d$ centered on the $\tilde z$ axis  in the region $|\tilde x
| < \tilde w$ and $|\tilde y| < d/2$, where $d \ll \tilde w$ [see figure 1].  
%***** Fig.1 ************************
\begin{figure}
\begin{center}
\includegraphics[width=8cm]{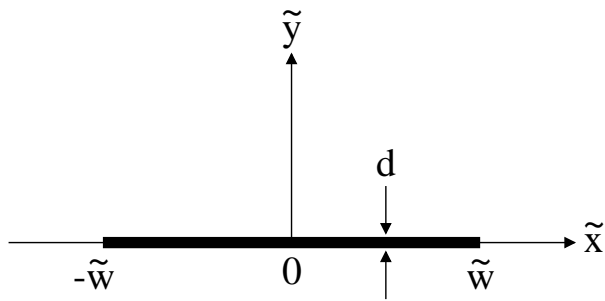}
\end{center}
\caption{Long thin superconducting strip of thickness $d$ and width $2 \tilde w$.}
\label{strip}
\end{figure}
(We
attach  tildes to all quantities related to the single strip.)
Since we are not concerned with details of how the
current density varies across the film thickness, we consider only
the sheet-current density
$\tilde {\bm K}(\tilde x) =
\int_{-d/2}^{d/2}\tilde {\bm j} d\tilde y =
\hat z
\tilde K_z(\tilde x)$.
If the film thickness
$d$ is less than the London penetration depth $\lambda$, we assume that the
two-dimensional screening length (or Pearl length) \cite{Pearl64}
$\Lambda = 2\lambda^2/d$ is much smaller than $\tilde w$.  We consider the general
case when the strip carries a current in the $\tilde z$ direction and  is
subjected to a perpendicular applied field $\tilde H_a$ in the
$\hat y$ direction. Outside the strip, the net magnetic field $\tilde {\bm
H}(\tilde x,\tilde y) = \hat x \tilde H_x(\tilde x,\tilde y) + \hat y
\tilde H_y(\tilde x,\tilde y)$ obeys  $\tilde \nabla \times \tilde {\bm
H}=0$ and 
$\tilde \nabla \cdot \tilde {\bm H}=0$.

For a two-dimensional problem such  as this, it is convenient to introduce
a complex magnetic field $\tilde H(\tilde \zeta) = \tilde H_y(\tilde x,
\tilde y) +i \tilde H_x(\tilde x, \tilde y)$, which  is  an
analytic function of the complex variable
$\tilde
\zeta =
\tilde x + i \tilde y$ outside the strip.  
The real and imaginary parts of $\tilde H(\tilde \zeta)$ obey the
Cauchy relations, $\partial \tilde H_y/\partial \tilde x = \partial
\tilde H_x/\partial \tilde y$ and $\partial \tilde H_y/\partial \tilde y =
-\partial \tilde H_x/\partial \tilde x$, which guarantee that $\tilde {\bm
H}$ obeys  
$\tilde \nabla \times \tilde {\bm H} = 0$ and $\tilde \nabla \cdot
\tilde {\bm H} = 0$, respectively.  The Biot-Savart law can be expressed as 
\begin{equation}
\tilde H(\tilde \zeta) = \tilde H_a
+\frac{1}{2\pi}\int_{-\tilde w}^{\tilde w}\frac{\tilde K_z(\tilde u)
d\tilde u}{\tilde \zeta - \tilde u}.
\label{Htilde}
\end{equation}
Using the property that $1/(x\pm i \epsilon) = (P/x) \mp i \delta(x)$,
where $\epsilon$ is a positive infinitessimal and  $P$ denotes the principal
value, we obtain 
\begin{equation}
\tilde H_y(\tilde x, 0) = \tilde H_a
+\frac{P}{2\pi}\int_{-\tilde w}^{\tilde w}\frac{\tilde K_z(\tilde u)
d\tilde u}{\tilde x - \tilde u}
\end{equation}
and 
\begin{eqnarray}
\tilde H_x(\tilde x, \pm \epsilon)&=& \mp \tilde K_z(\tilde x)/2, ~~|\tilde
x| < \tilde w, \nonumber \\
&=& 0, ~~{\rm otherwise}.
\end{eqnarray}
The net current carried in the $\tilde z$ direction is
\begin{equation}
\tilde I_z = \int_{-\tilde w}^{\tilde w} \tilde K_z(\tilde x) d\tilde x.
\label{Iz0tilde}
\end{equation}
Note from  (\ref{Htilde}) that an expansion of $\tilde H(\tilde \zeta)$
in powers of $1/\tilde \zeta$ yields \cite{Mawatari03}
$\tilde H(\tilde
\zeta) =
\tilde H_a + \tilde I_z/2\pi \tilde \zeta + {\rm O}(1/\tilde
\zeta^2)$.

Solutions for the magnetic-field and current-density
distributions for a long thin strip are known for many different physical
situations.
It has been shown by Mawatari in  \cite{Mawatari96} how  
known solutions $\tilde H(\tilde \zeta)$ and $\tilde K_z(\tilde x)$ for a
single isolated strip can be used to  generate the corresponding solutions
$H(\zeta)$ and $K_z(x)$ for an $X$ array, i.e., an
infinite periodic array of identical coplanar strips of width $2w$ and
periodicity
$L$ in the plane $y = 0$, as sketched in figure 2.
%***** Fig.2 ************************
\begin{figure}
\begin{center}
\includegraphics[width=8cm]{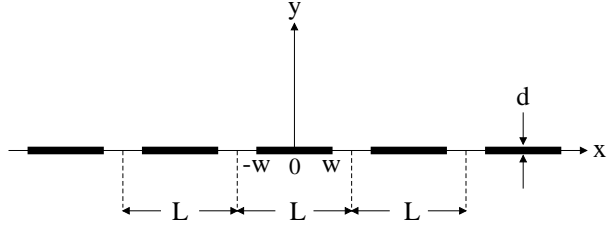}
\end{center}
\caption{X array of long thin superconducting strips of thickness $d$, width $2
w$, periodicity $L$, and separation $\Delta = L -2w$.}
\label{Xarray}
\end{figure}
The complex field $H(\zeta) = H_y(x,y) + i H_x(x,y)$, which is
an analytic function of the complex variable
$\zeta = x+iy$ outside the strips, describes the components of the net
magnetic field ${\bm H}(x,y) = \hat x H_x(x,y) + \hat y
H_y(x,y)$  produced by a sheet-current density $\bm K(x) = \hat z K_z(x)$
and a magnetic field $H_a$ applied in the $y$ direction.  The
periodicity requires that the magnetic field and sheet-current density obey
${\bm H}(x+L,y) ={\bm H}(x,y)$ and $\bm K(x+L) =\bm K(x)$. 

Corresponding solutions for the periodic complex field $H(\zeta)$ can be obtained
from $\tilde H(\tilde \zeta)$ using the conformal mapping \cite{Mawatari96}
\begin{equation}
\tilde \zeta = (L/\pi) \tan(\pi \zeta/L),
\label{tzeta}
\end{equation}
or its inverse
\begin{equation}
 \zeta = (L/\pi) \arctan(\pi \tilde \zeta/L).
\label{zeta}
\end{equation}
These equations also describe the relations between $\tilde x$ and $x$, $\tilde w$
and $w$, or $\tilde u$ and $u$, which will appear in later expressions.
Then $H(\zeta) = \tilde H(\tilde \zeta)$, $K_z(x) = \tilde K_z(\tilde x)$, and 
the  Biot-Savart law yields \cite{Mawatari96}
\begin{equation}
H(\zeta) = H_a
+\frac{1}{2L}\int_{-w}^{w} K_z(u)\cot\Big[\frac{\pi(\zeta-u)}{L}\Big]du,
\label{cot}
\end{equation}
where the magnetic field applied in the $y$ direction in the $\zeta$ plane is 
\cite{Mawatari96}
\begin{equation}
H_a = \tilde H_a
-\frac{1}{2\pi}\int_{-\tilde w}^{\tilde w}
\frac{\tilde K_z(\tilde u) \tilde u d\tilde u}{(L/\pi)^2+\tilde u^2}
\label{HaHatilde}
\end{equation}
and the current carried in the $z$ direction by one of the strips shown in
figure \ref{Xarray} is 
\begin{equation}
I_z = \int_{-w}^w K_z(x) dx.
\label{Iz0}
\end{equation}
Note by comparing  (\ref{Iz0tilde}) and (\ref{Iz0}) that
$I_z$ is in general not the same as $\tilde I_z$.
The relations (\ref{HaHatilde}) and (\ref{Iz0}) can be obtained from 
 (\ref{cot}) by
applying the requirement that 
\begin{equation}
H(\pm i \infty) = \tilde H(\pm i L/\pi) = H_a \mp i I_z/2L.
\label{lims}
\end{equation}
Along the $x$ axis, we have the properties
$H(x\pm i
\epsilon) = \tilde H(\tilde x \pm i \epsilon)$, $H_x(x\pm i
\epsilon) = \mp K_z(x)/2$, and 
\begin{equation}
H_y(\pm L/2,0) =
\tilde H_a
=H_a +
\frac{1}{2L} \int_{-w}^{w}
K_z(u) \tan(\frac{\pi u}{L}) du.
\label{HatildeHa}
\end{equation}  

In the following sections, we shall obtain $H(\zeta)$  using $H(\zeta) =
\tilde H(\tilde \zeta)$,  $K_z(x) = \tilde K_z(\tilde x)$,  (\ref{tzeta}) and
(\ref{lims}), and various trigonometric identities.

\section{\label{bulk-pinning}%
Critical current due only to bulk pinning}

When the critical current of  a single superconducting strip
is due solely to bulk pinning, characterized by the bulk-pinning critical
current density 
$J_p$ (here assumed to be field-independent, an assumption to be justified in
Sec.\ \ref{conclusions}),  the corresponding critical sheet-current density is
$K_p = J_p d$, and
$\tilde K_z = K_p$ for $|\tilde x| < \tilde w$, such that the critical
current is 
$\tilde I_p = 2 K_p \tilde w$ and the complex field is
\begin{equation}
\tilde H(\tilde \zeta) = \tilde H_a
+\frac{K_p}{2\pi}\ln\Big(\frac{\tilde \zeta + \tilde w}{\tilde \zeta -
\tilde w}\Big).
\end{equation}

The corresponding complex field surrounding the $X$ array shown in figure
\ref{Xarray} is
\begin{equation}
H(\zeta) = H_a
+\frac{K_p}{2\pi}
\ln\Big[\frac{\sin{[\pi(\zeta+w)/L]}}{\sin{[\pi(\zeta-w)/L]}}\Big],
\label{HKp}
\end{equation}
where $H_a=\tilde H_a=H_y(\pm L/2,0)$.
Since $K_z = \tilde K_z = K_p$, we immediately find for one of the strips
in the $X$ array that its critical current  is
$I_p = 2K_p w$, its average critical sheet-current density is
$K_p$, and its average critical
current density  is
$J_p$.
Note that when $w < L/2$, the self-field contribution to 
$H_y(x,0)$ is positive and diverges logarithmically when $x
\rightarrow w$; it is negative and has a similar
divergence when $x
\rightarrow -w$. In the limit that
$w
\rightarrow L/2$, however, we recover the complex potential generated by an
infinite film  carrying a sheet  current $K_z = K_p$ in a perpendicular
applied field
$H_a$, 
\begin{equation}
H(\zeta) = H_a
\mp i K_p/2,
\label{HKpinf}
\end{equation}
where the upper (lower) sign holds when $y > 0$ ($y<0$).

\section{\label{edge_barrier_only}%
Critical current due only to edge barriers}

Consider a single bulk-pinning-free superconducting strip in which the
critical current is determined by a geometrical edge barrier (or surface
barrier at the edge). As discussed in  \cite{Benkraouda96}, the
current and field distributions at the critical current have two possible
forms, depending upon the strength of the applied field $\tilde H_a$.
For small values of
$\tilde H_a$, the distributions are simply the Meissner response to
the applied field and current, but for larger values of $\tilde H_a$,
there is a domelike magnetic field distribution inside the strip.

The complex magnetic field describing the Meissner response
to a magnetic field
$\tilde H_a$ applied in the $\tilde y$ direction and to a net current
$\tilde I$ applied in the
$z$ direction is \cite{Benkraouda96} 
\begin{equation}
\tilde H(\tilde \zeta) = \frac{\tilde H_a \tilde \zeta +\tilde
I/2\pi}{(\tilde \zeta^2-\tilde w^2)^{1/2}},
\end{equation}
and the sheet-current density in the strip $(|\tilde x|<\tilde w)$
is 
\begin{equation}
\tilde K_z(\tilde x) = \frac{2 \tilde H_a \tilde x +\tilde
I/\pi}{(\tilde w^2 -\tilde x^2)^{1/2}}.
\label{KzMeissner}
\end{equation}

If the strip contains a domelike magnetic-flux distribution in the region
$\tilde a < \tilde x < \tilde b$, where the sheet-current density is
zero, the complex magnetic field is \cite{Benkraouda96}
\begin{equation}
\tilde H(\tilde \zeta) =\tilde H_a 
\Big[\frac{ (\tilde \zeta-\tilde a) ( \tilde \zeta - \tilde
b)}{\tilde
\zeta^2-\tilde w^2}\Big]^{1/2},
\end{equation}
the sheet-current density in the strip $(|\tilde x|<\tilde w)$ is 
\begin{eqnarray}
\tilde K_z(\tilde x)& =&
 2 \tilde H_a \sqrt{\frac{(\tilde x-\tilde a) ( \tilde x -
\tilde b)}{\tilde w^2 -\tilde x^2}}, ~~
\tilde b < \tilde x < \tilde w,
\nonumber \\
& =&
 -2 \tilde H_a \sqrt{\frac{(\tilde a-\tilde x) ( \tilde b -
\tilde x)}{\tilde w^2 -\tilde x^2}}, 
~~-\tilde w < \tilde x < \tilde a,
\nonumber \\
& =&
0, ~~{\rm otherwise},
\end{eqnarray}
and the net current in the $\tilde z$ direction is \cite{Benkraouda96}
\begin{equation}
\tilde I = -\pi (\tilde a + \tilde b)\tilde H_a.
\end{equation}

The corresponding quantities for the $X$ array are, for the Meissner
response to an applied field $H_a$ in the $y$ direction and an
applied current $I$ in the $z$ direction, the complex field
\begin{equation}
H(\zeta) = \frac{H_a \sin(\frac{\pi \zeta}{L}) +(I/2L)\cos(\frac{\pi
\zeta}{L})}{[\sin\!\big(\frac{\pi(\zeta+w)}{L}\big)
\sin\!\big(\frac{\pi(\zeta-w)}{L}\big)]^{1/2}}
\label{HI}
\end{equation}
and the sheet-current density in the strip $(|x|<w)$  
\begin{equation}
K_z(x) = \frac{2 H_a \sin(\frac{\pi x}{L}) +(I/L) \cos(\frac{\pi
x}{L})} {\sqrt{\sin\!\big(\frac{\pi(w+x)}{L}\big)
\sin\!\big(\frac{\pi(w-x)}{L}\big)}},
\label{KzHI}
\end{equation}
where $\tilde H_a=H_a/\cos(\pi w/L) = H_y(\pm L/2,0)$ and
$\tilde I =  I/\cos(\pi w/L)$.

If all the strips in the $X$ array contain domelike magnetic-flux
distributions identical to the one in the region $a < x < b$,  the complex
field is
\begin{equation}
 H( \zeta) =\frac{H_a}{\cos\!\big(\frac{\pi (a+b)}{2L}\big)} 
\bigg[\frac{
\sin\!\big(\frac{\pi(\zeta-a)}{L}\big)
\sin\!\big(\frac{\pi(\zeta-b)}{L}\big)}
{\sin\!\big(\frac{\pi(\zeta+w)}{L}\big)
\sin\!\big(\frac{\pi(\zeta-w)}{L}\big)}\bigg]^{1/2},
\label{Hdome}
\end{equation}
 the sheet-current density in the central strip $(|x|<w)$ is 
\begin{eqnarray}
K_z(x)& =&
 \frac{2H_a}{\cos\!\big(\frac{\pi (a+b)}{2L}\big)} 
\sqrt{\frac{
\sin\!\big(\frac{\pi(x-a)}{L}\big)
\sin\!\big(\frac{\pi(x-b)}{L}\big)}
{\sin\!\big(\frac{\pi(w+x)}{L}\big)
\sin\!\big(\frac{\pi(w-x)}{L}\big)}}, 
\nonumber \\
&&~~~~~~~~~~~~~~~~~~~~~~~~~~~~~~~~~~~~~~b < x < w,
\nonumber \\
& =&
 -\frac{2H_a}{\cos\!\big(\frac{\pi (a+b)}{2L}\big)} 
\sqrt{\frac{
\sin\!\big(\frac{\pi(a-x)}{L}\big)
\sin\!\big(\frac{\pi(b-x)}{L}\big)}
{\sin\!\big(\frac{\pi(w+x)}{L}\big)
\sin\!\big(\frac{\pi(w-x)}{L}\big)}}, 
\nonumber \\
&&~~~~~~~~~~~~~~~~~~~~~~~~~~~~~~~~~~~~- w < x <  a,
\nonumber \\
& =&
0, ~~{\rm otherwise},
\label{Kzdome}
\end{eqnarray}
where 
\begin{equation}
\tilde H_a=  \frac{H_a\sqrt{\cos\!\big(\frac{\pi
a}{L}\big)\cos\!\big(\frac{\pi
b}{L}\big)}}{\cos\!\big(\frac{\pi
(a+b)}{2L}\big)\cos\!\big(\frac{\pi
w}{L}\big)} =H_y(\pm L/2,0),
\end{equation}
and the net current in the $z$ direction in the strip $|x| < w$ is
\begin{equation}
I = -2L H_a \tan [\pi
(a+b)/2L].
\label{Idome}
\end{equation}

The divergences in the above expressions for $H$ and $K_z$ at $x = \pm w$ are
artifacts of ignoring the finite thickness $d$ of the film.  It is well known
that these divergences are cut off at a length scale $\Lambda_c$, the larger of  
$\Lambda = 2 \lambda^2/d$ and $d/2$ [i.e., $\Lambda_c =
\max(\Lambda,d/2)$].  
To determine the critical current of one of the strips of the $X$
array, we use  the approximations applied in 
\cite{Elistratov02} and \cite{Zeldov94}.  
Accordingly, we  estimate the
the sheet-current density at an edge by evaluating the diverging
inverse square root in the expression for
$K_z$ at a distance
$\Lambda_c$ from that edge; for example, for the edge at $x=w$, in 
(\ref{KzHI}) and (\ref{Kzdome}) we replace
$x$ by $w$ in the numerators and by 
$x_c = w-\Lambda_c$ in the denominators and use the fact that the cut-off length
scale obeys 
$\Lambda_c  \ll
w$. To account for the edge barrier, we assume that vortices nucleate and enter the
superconductor when the magnitude of
$K_z$ at either edge of the strip reaches the critical value $K_s = j_s d$
at which the barrier is overcome. For an ideal edge, $j_s$ is equal to the
Ginzburg-Landau depairing current density $j_{GL}$ \cite{Aslamazov,VMB},
but for an extremely defected edge, $j_s$ may become very small.
Applying this procedure to  (\ref{KzHI}) for $H_a =0$, we
obtain the following approximation to the zero-field surface-barrier critical
current for any one of the strips in the $X$ array, 
\begin{equation}
I_s(0) \equiv I_{s0} = K_s(2\pi \Lambda_c L \tan \theta)^{1/2}  ,
\label{Is00}
\end{equation}
where $\theta = \pi w /L$.
Alternatively,  by estimating the local
magnetic field at the edge of the strip \cite{Benkraouda96,Zeldov94},
one obtains  (\ref{Is00}) but with $K_s \approx 2 H_s$, where $H_s$ is the
critical field at which the barrier is overcome.  We expect that $H_s$ is at least
as large as the lower critical field $H_{c1}$, but under favorable circumstances it
may approach the bulk thermodynamic critical field $H_c$.

As a function of the applied field $H_a$, the surface-barrier
critical current is found from   (\ref{KzHI}) to obey
\begin{equation}
I_s(H_a)/I_s(0) = 1-h
\label{IsH0} 
\end{equation}
for small $h$, where 
\begin{equation}
h = H_a/(I_{s0}/2L\tan \theta).
\label{h}
\end{equation}  
When $L \rightarrow \infty$,  (\ref{Is00}) and (\ref{IsH0}) reduce to
corresponding results found in 
\cite{Benkraouda96} and \cite{Elistratov02} in low fields for
isolated bulk-pinning-free strips.

In the linear region of $I_s(H_a)$ given in  (\ref{IsH0}), the flux
flow producing the voltage at a current just above $I_s$ can be described
as the nucleation of vortices at $x = w$ as they overcome the
geometrical edge barrier, followed by rapid motion  across the
strip and annihilation at $x=-w$.  This can occur only when $K_z(x) > 0$ for
all $|x| < w$. For increasing $H_a$, however, as can be seen from 
(\ref{KzHI}),
$K_z(-w)$  becomes zero at the critical current $I_s(H_a)$ when $H_a =
I_s(H_a)/2L\tan \theta$. Combining this condition with 
(\ref{IsH0}), we find that this occurs when $H_a = H_{d0}$, where
\begin{equation}
H_{d0} =I_{s0}/4L\tan \theta,
\end{equation}
and the subscripts ($d0$) denote the onset of a dome for zero bulk
pinning.  Note that we also may write $h = H_a/2H_{d0}$.  The linear
portion of
$I_s(H_a)$ vs
$H_a$ given in  (\ref{IsH0}) ends when $H_a = H_{d0}$ or
$h=h_{d0}\equiv 1/2$. 

For applied fields $H_a > H_{d0}$ (or $h > h_{d0}=1/2$), each strip
contains a domelike magnetic field distribution at the critical current,
and $I_s(H_a)$ is no longer a linear function of $H_a$.  The flux
flow producing the voltage at a current just above $I_s$ can be described
as the nucleation of vortices at $x = w$, as they overcome the
geometrical edge barrier, followed by rapid motion from $x = w$ to $x = b$,
the right boundary of the magnetic field dome.  The vortices inside
the dome move collectively very slowly, and vortices at $x = a \approx -w$
get pushed out of the dome and annihilate at 
$x=-w$. We therefore calculate the critical current by setting $a = -w$ in
 (\ref{Kzdome}), evaluating $K_z$ at $x = w$ using the above
approximation method, and setting it equal to $K_s$.  Using  (\ref{Is00}) and
(\ref{h}) we then find that the value of $b$ at the
critical current is
$b_c$, where 
\begin{equation}
\tan[\pi (w-b_c)/2L] = \tan(\pi w/L)/4h^2.
\label{tandome}
\end{equation}
(Note that $b_c = -w$ when $h$ = 1/2.)
Substituting this into  (\ref{Idome}), we obtain $I_s(H_a)$ for  
$H_a \ge
H_{d0}$ (or $h \ge h_{d0} = 1/2$).  The critical
current  of  one of the strips of the
$X$ array is thus given by 
\begin{eqnarray}
\frac{I_s(H_a)}{I_{s0}}&=& 1-h, ~~0 \le h \le h_{d0} =\frac{1}{2}, \nonumber
\\ &=&\frac{1}{4h}, ~~h_{d0} =
\frac{1}{2} \le h < h_{max},
\end{eqnarray}
when there is no bulk pinning and the critical current is determined only by a
geometrical edge barrier (surface barrier).
We expect our theory to be accurate only at sufficiently low
applied fields that $h < h_{max}$ or $H_a < H_{max}$, where $h_{max}$ is the
reduced field at which
$b_c = w-\Lambda_c$.  From   (\ref{tandome}) we obtain $h_{max} =
[(L/2\pi\Lambda_c)\tan \theta]^{1/2} \geq (w/2\Lambda_c)^{1/2} \gg 1$ and $H_{max}
\approx K_s/2$. 

\section{\label{both_effects}%
Critical current due to both edge barriers and bulk pinning}

In this section we calculate the complex fields and critical current $I_c$
of one of the strips in the $X$ array when the critical current is due not
only to  a geometrical edge barrier but also bulk pinning, characterized by
a bulk-pinning field-independent critical sheet-current density $K_p$. An
important parameter is the ratio\footnote{In the limit $L \rightarrow \infty$, 
$p = I_p/I_{s0}$ as
in  \cite{Elistratov02}.} 
\begin{equation}
p = \frac{I_p}{I_{s0}} \frac{\sin \theta}{\theta},
\label{p}
\end{equation} 
where $I_p = 2K_pw$ [see
Sec.\
\ref{bulk-pinning}] is the critical current due to bulk pinning in the
absence of an edge barrier,  $I_{s0}$ [see Sec.\ \ref{edge_barrier_only}]
is the critical current due to the edge barrier in the absence of bulk
pinning when $H_a = 0$, and $\theta = \pi w/L$.  

\subsection{Region I: One vortex-free zone with no domes}

We first consider the case of
relatively weak bulk pinning when 
$p < 2/\pi$.  In low fields ($0< H_a < H_d$ or $ 0 < h < h_d$, region I of
figure 3),
%***** Fig.3 ************************
\begin{figure}
\begin{center}
\includegraphics[width=8cm]{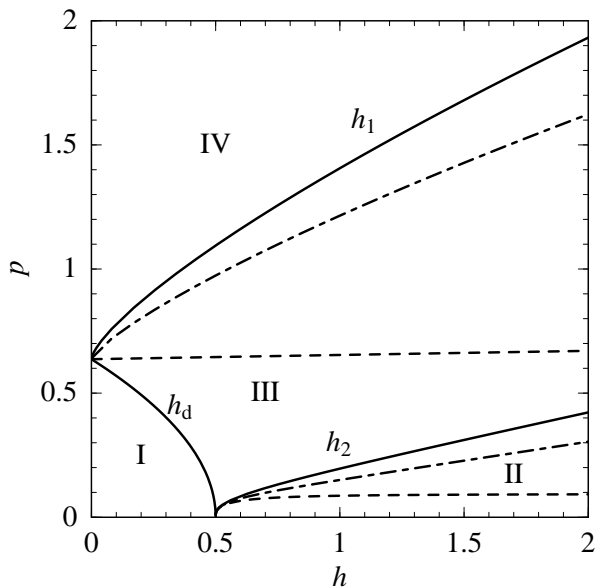}
\end{center}
\caption{Diagram showing the behavior at the critical current vs reduced field $h$
and bulk pinning parameter $p$.  In
region I, the strip is vortex-free [$H_y(x,0)=0$].  In region II,
there are two zones:  a vortex dome ($-w < x <
b$) and a vortex-free zone ($b < x < w$). In region III, there are three zones:  a
vortex dome ($a < x < b$) between two vortex-free zones ($-w < x < a$ and
$b < x < w$).  In region IV, there
are four zones: an antivortex dome
($a < x < c$) where
$H_y(x,0)<0$,   a vortex dome ($c < x < b$) where $H_y(x,0)>0$, and
two vortex-free zones ($-w < x < a$ and $b < x < w$).  In each vortex-free zone
$K_z(x)>K_p$, and  under each vortex dome
$K_z = K_c$. The curve
$h_d(p)$ separates regions I and III,  the curves
$h_1(p,\theta)$  separate regions III and IV, and the curves
$h_2(p,\theta)$ separate regions II and III.  These curves are shown for
$\theta = \pi w/L = 0$ (solid),
$\theta = 0.5
\pi/2$ (dot-dashed), and  $\theta = 0.99 \pi/2$ (dashed).}
\label{pvsh}
\end{figure}
vortices nucleate at the right edge of the strip ($x=w$) when
$I_z$ slightly exceeds $I_s(H_a)$. When $K_z(x)$ in  (\ref{KzHI}) is
greater than
$K_p$ for all $x$ in the strip ($|x| < w$), nucleating vortices are driven
entirely across the strip to the opposite side at $x = -w$, where they
annihilate. The critical current is then  $I_c(H_a,p,\theta) = I_s(H_a)$, and
the normalized critical current is 
\begin{equation}
i_c(h,p,\theta) = I_c(H_a,p,\theta)/I_{s0} = 1-h,
\label{icI}
\end{equation}
the same result as in  (\ref{IsH0}).  Thus if  $p$ is small, the
critical current $I_c(H_a,p,\theta)$ for small $H_a$ is still completely
dominated by the edge barrier and is independent of the strength of the bulk
pinning. From   (\ref{KzHI}) we see that $K_z(x)$ vs $x$ has a minimum,
$K_z(x_{\rm min})$, at $x = x_{\rm min}$, where 
\begin{equation}
\tan(\pi x_{\rm min}/L) = -(2LH_a/I)\tan^2(\pi w/L).
\end{equation}
The minimum deepens [i.e., $K_z(x_{\rm min})$ decreases] with increasing
$H_a$, and
$K_z(x_{\rm min}) = K_p$ when $I = I_c(H_a,p,\theta)$ and $H_a = H_d$ or $h =
h_d$, such that 
\begin{equation}
\tan(\pi x_{\rm min}/L) = -[h_d/(1-h_d)]\tan \theta,
\end{equation}
where 
\begin{equation}
h_d = \frac{H_d}{I_{s0}/2L\tan \theta} = \frac{1}{2}[1-(\frac{ \pi
p}{2})^2].
\end{equation}
As a result, for small $p<2/\pi$, the linear behavior of $I_c(H_a)$ vs
$H_a$, given by 
$i_c(h,p,\theta)$ vs
$h$ in  (\ref{icI}), holds only for $0 \le h \le h_d$.

When $p<2/\pi$ and $h>h_d$,  or when $p > 2/\pi$ and $h$ has any value,
i.e., for $h$ and $p$ outside region I of figure 3, domelike magnetic-field
distributions occur at the critical current $I_c(H_a)$. Using the X-array
method \cite{Mawatari96}, we can obtain the complex field
$H(\zeta)$ and associated sheet-current density $K_z(x)$ in one of the
strips of the $X$ array from the corresponding solution for an isolated strip 
\cite{Elistratov02}.  For the $X$ array, we find $H(\zeta) = \tilde
H(\tilde \zeta)$, where 
\begin{equation} 
\tilde H(\tilde \zeta) =\tilde P(\tilde a, \tilde b, \tilde \zeta)
[\tilde H_a +
\frac{K_p}{2\pi}\tilde Q(\tilde a,\tilde b,\tilde \zeta)],
\label{HzetaI}
\end{equation} 
where
\begin{equation}
\tilde P(\tilde a, \tilde b, \tilde \zeta) =
\Big[\frac{(\tilde \zeta-\tilde
a)(\tilde \zeta-\tilde b)}{(\tilde \zeta^2-\tilde
w^2)}\Big]^{1/2},
\label{tildeP}
\end{equation}
\begin{eqnarray} 
\tilde Q(\tilde a,\tilde b,\tilde \zeta) &=&
\int_{\tilde a}^{\tilde
b}\frac{\sqrt{\tilde w^2-\tilde u^2}}{(\tilde
\zeta-\tilde u)\sqrt{(\tilde u-\tilde a)(\tilde b-\tilde u)}}d\tilde u 
\label{tildeQ} \\ &=&
\frac{2(\tilde w+\tilde a)}{\sqrt{(\tilde
 w-\tilde a)(\tilde  w+\tilde b)}}
\Big[{\mathbf{\Pi}}
\Big(\frac{\tilde b-\tilde a}{\tilde w+\tilde
b},\tilde q\Big) \nonumber \\
&&-\frac{(\tilde \zeta-\tilde
w)}{(\tilde
\zeta-\tilde a)}
{\mathbf{\Pi}}\Big(\frac{(\tilde \zeta+\tilde
w)(\tilde b-\tilde a)}{(\tilde \zeta-\tilde a)(\tilde w+\tilde
b)},\tilde q\Big)\Big] \label{tildeQa} \\
&=&
-\frac{2(\tilde w-\tilde b)}{\sqrt{(\tilde
 w-\tilde a)(\tilde  w+\tilde b)}}\Big[{\mathbf{\Pi}}
\Big(\frac{\tilde b-\tilde a}{\tilde w-\tilde
a},\tilde q\Big) \nonumber \\
&&-\frac{(\tilde \zeta+\tilde
w)}{(\tilde
\zeta-\tilde b)}
{\mathbf{\Pi}}\Big(\frac{(\tilde \zeta-\tilde
w)(\tilde b-\tilde a)}{(\tilde \zeta-\tilde b)(\tilde w-\tilde
a)},\tilde q\Big)\Big], 
\label{tildeQb}
\end{eqnarray}
and 
\begin{equation} \tilde q^2 = \frac{2\tilde w(\tilde b-\tilde
a)}{(\tilde w-\tilde a)(\tilde w+\tilde b)}.
\label{tildeq}
\end{equation} 
The integral
in  (\ref{tildeQ}) is expressed in terms of complete elliptic integrals 
of the third kind \cite{Grad00,Abram67,Math,Selfridge58,Byrd}
${\mathbf{\Pi}}(n,k)$, where
$n$ is called either the characteristic or the parameter, and $k$ is called
the modulus.  The mappings of  (\ref{tzeta}) and (\ref{zeta}) define the
relations between $\tilde \zeta$ and $\zeta$,  $\tilde a$ and $a$, 
$\tilde b$ and $b$, or  $\tilde w$ and $w$, and also guarantee that $H(\zeta)$ is
periodic in the $x$ direction with period $L$.  The applied magnetic field $H_a$ in
the
$y$ direction  and the current $I_z$ carried in the $z$
direction by one of the strips shown in figure \ref{Xarray}, obtained from 
(\ref{lims}), obey  
\begin{equation}
H_a = {\rm Re}\{\tilde P(\tilde a, \tilde b, i L/\pi)
[\tilde H_a +
\frac{K_p}{2\pi}\tilde Q(\tilde a,\tilde b,i L/\pi)]\}
\label{Hadomes}
\end{equation}
and 
\begin{equation}
I_z = -2L{\rm Im}\{\tilde P(\tilde a, \tilde b, i L/\pi)
[\tilde H_a +
\frac{K_p}{2\pi}\tilde Q(\tilde a,\tilde b,i L/\pi)]\},
\label{Izdomes}
\end{equation}
where
\begin{equation}
\tilde P(\tilde a, \tilde b, i L/\pi) = 
\frac{\cos \theta}{\sqrt{\cos(\pi a/L)\cos(\pi b/L)}}e^{i\pi(a+b)/2L}
\end{equation}
and $\theta = \pi w /L.$

The quantities of primary interest to us are $H_y(x,0) = {\rm Re}H(x)$ and $K_z(x)
= -2 {\rm Im} H(x+i\epsilon)$,
\begin{eqnarray} 
H_y(x,0) &= &{\rm Re} \tilde H(\tilde x) \\
&=& \tilde P_0(\tilde a, \tilde b, \tilde x)[\tilde H_a 
+\frac{K_p}{2 \pi} {\rm Re} \tilde Q(\tilde a, \tilde b, \tilde x)],
\label{Hy} \\
&&\;\;\;\;\;\;\;\tilde a < \tilde x < \tilde b  \;{\rm or}\; |\tilde x| > \tilde w,
\nonumber \\ 
&=& 0, -\tilde w < \tilde x < \tilde a \;{\rm or}\;
 \tilde b < \tilde x < \tilde w.
\end{eqnarray}
\begin{eqnarray} 
K_z(x) & = & \tilde K_z(\tilde x) = -2{\rm Im} \tilde H(\tilde x+i\epsilon) \\
&=& -2 \tilde P_0(\tilde a, \tilde b, \tilde x)[\tilde H_a 
+\frac{K_p}{2 \pi}\tilde Q(\tilde a, \tilde b, \tilde x)],
\label{Kzleft} \\
&& \;\;\;\; -\tilde w < \tilde x < \tilde a,  \nonumber  \\
&=& K_p, \;\tilde a < \tilde x < \tilde b,
\label{Kzmiddle} \\
&=& +2 \tilde P_0(\tilde a, \tilde b, \tilde x)[\tilde H_a 
+\frac{K_p}{2 \pi}\tilde Q(\tilde a, \tilde b, \tilde x)], 
\label{Kzright} \\ 
&&\;\;\;\;\;\;\;\; \tilde b < \tilde x < \tilde w, \nonumber
\end{eqnarray}
where 
\begin{equation}
\tilde P_0(\tilde a, \tilde b, \tilde x) 
= \sqrt{\frac{|\tilde x - \tilde a||\tilde x - \tilde b|}{|\tilde x^2 - \tilde
w^2|}}.
\label{tildeP0}
\end{equation}
For the calculation of $H_y(x,0)$ for $a < x < b$ (i.e., $\tilde a < \tilde x <
\tilde b$), taking the real part of $\tilde Q$ in  (\ref{Hy})
corresponds to taking the principal value of the integral of  (\ref{tildeQ}).

\subsection{Region IV: Two vortex-free zones, one vortex dome, and one antivortex
dome}

For values of $p > 2/\pi$ [see (\ref{p})], and small values of $h$, i.e., for
$h$ and $p$ in region IV of figure \ref{pvsh}, the vortex distribution in  each of
the strips at the critical current can be described as a double dome, consisting
of a vortex dome adjacent to an antivortex dome.  Just above the
critical current, vortices nucleate at $x = w$, where
$K_z(x_c) = K_s$,  move rapidly to the left through an otherwise vortex-free
region ($b < x < w$), and then move slowly to the left through a vortex-filled
region (the vortex dome), $c < x < b$.   Antivortices nucleate at $x = -w$,
where $K_z(-x_c) = K_s$, move rapidly to the right through an otherwise
vortex-free region ($-w < x < a$), and then move slowly to the right through an
antivortex-filled region (the antivortex dome),
$a < x < c$.   Vortices and antivortices annihilate at $x = c$, where the two domes
meet and $H_y(c,0) = 0$.

\begin{figure}
\begin{center}
\includegraphics[width=8cm]{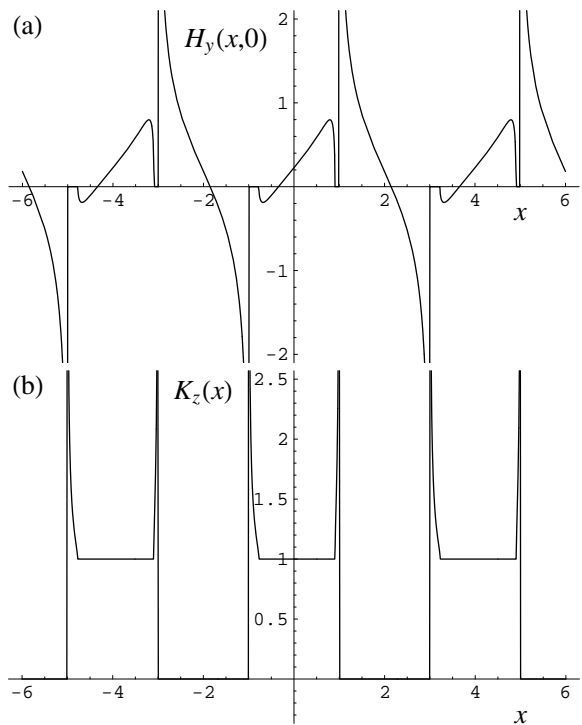}
\end{center}
\caption{Plots of (a) $H_y(x,0)$ (in units of $I_{s0}/2L\tan \theta$) and (b) 
$K_z(x)$ (in units of $K_p$) vs $x$ (in units of
$w$) for $h = 0.2$, $p = 0.85$, and $\theta = \pi w /L = 0.5 \pi/2$ (region IV),
for which $i_c = I_c/I_{s0} = 1.11$ and $I_c/I_p = 1.18$,
showing three strips ($-5 < x/w < -3,$
$-1 < x/w < 1,$ and $3 < x/w < 5$) and the gaps around them ($-7 < x/w < -5$, -$3
< x/w < -1$, $1 < x/w < 3,$ and $ 5 < x/w < 7)$.}
\label{hyIV}
\end{figure}

Four equations must be solved simultaneously for
$a$, $b$, $\tilde H_a$, and the critical current
$I_c(H_a)$ for known values of $h$, $p$, and $\theta = \pi w/L$ in region IV of
figure
\ref{pvsh}. One condition is that
$K_z(w) = K_s$, 
which we evaluate by replacing $\tilde x$  in the denominator of 
(\ref{tildeP0}) by $\tilde x_c$, where $x_c  = w -
\Lambda_c$ and $\tilde x_c = (L/\pi) \tan(\pi x_c/L) \approx \tilde w -
\Lambda_c \sec^2\theta $, and by replacing $\tilde x$ by 
$\tilde w$ in
$\tilde Q(\tilde a, \tilde b, \tilde x)$ (\ref{tildeQa}) and in the
numerator of  (\ref{tildeP0}).
This  yields from   (\ref{p}) and (\ref{Kzright}) 
\begin{equation}
\sqrt{(1-\tilde a')(1-\tilde b')}\tilde h \cos \theta
+(1+\tilde a')\sqrt{\frac{1-\tilde b'}{1+\tilde
b'}}{\mathbf{\Pi}}(\frac{\tilde b'-\tilde a'}{1+\tilde b'},\tilde q)p
=1,
\label{K+w}
\end{equation} where we use the normalized quantities $\tilde a'=\tilde a/\tilde w$
and
$\tilde b' = \tilde b/\tilde w$, as well as the definition
\begin{equation}
\tilde h = \tilde H_a/(I_{s0}/2L\tan\theta).
\label{tildeh}
\end{equation}
A second condition is that $K_z(-w) = K_s$, which we evaluate in a similar
manner with the help of    (\ref{p}), (\ref{tildeQb}), and (\ref{Kzleft}). 
The result is 
\begin{equation}
-\sqrt{(1+\tilde a')(1+\tilde b')}\tilde h \cos \theta
+(1-\tilde b')\sqrt{\frac{1+\tilde a'}{1-\tilde a'}}
{\mathbf{\Pi}}(\frac{\tilde b'-\tilde a'}{1-\tilde a'},\tilde q)p
=1.
\label{K-w}
\end{equation}
Equations (\ref{K+w}) and (\ref{K-w}) have nearly the same form as  (9)
and (10) in  \cite{Elistratov02} and reduce exactly to those equations
in the limit as $\theta = \pi w/L \to 0$ or $L \to \infty$. The third and fourth
equations  needed are obtained from  (\ref{Hadomes}) and (\ref{Izdomes}),
which we write in dimensionless form at $I_z = I_c$  as 
\begin{eqnarray}
h &=& \tilde P_1 \tilde h +(p/2) (\tilde P_1 \tilde Q_1
-\tilde P_2 \tilde Q_2) \sec\theta,
\label{hregionIV} \\
i_c &=& -\cot\theta[\tilde P_2 \tilde h +(p/2)  (\tilde P_1 \tilde Q_2
+\tilde P_2 \tilde Q_1)\sec\theta],
\label{icregionIV}
\end{eqnarray}
where 
the real quantities $ \tilde P_1,  \tilde P_2,  \tilde Q_1$ and
$
\tilde Q_2$ are defined via $\tilde P(\tilde a, \tilde b, iL/\pi)=\tilde P_1+ i
\tilde P_2$ and $\tilde Q(\tilde a, \tilde b, iL/\pi)=\tilde Q_1+ i \tilde
Q_2$, $h$, $\tilde h$, and $p$ are defined via 
(\ref{h}), (\ref{tildeh}), and (\ref{p}),  and $i_c(h,p,\theta) = I_c/I_{s0}$.
For given values of the strip width $2w$, periodicity length $L$, dimensionless
applied field
$h$, and dimensionless bulk pinning strength $p$, numerical solutions of 
(\ref{K+w}), (\ref{K-w}), (\ref{hregionIV}), and (\ref{icregionIV}) yield the
strip's dimensionless critical current $i_c(h,p,\theta)$ as well as the three
other unknowns, $\tilde a$ (or
$a$),
$\tilde b$ (or $b$), and $\tilde h$.
Shown in figure \ref{hyIV} (a) and (b) are sample plots of $H_y(x,0)$ and
$K_z(x)$ for $h$ and $p$ in region IV. 
%***** Fig.4 ************************

\subsection{h = 0}

Simplifications occur when calculating the self-field critical current when $p >
2/\pi$ and
$H_a$ = 0 ($h = 0$), for which $\tilde a = -\tilde b$ ($a = -b$) and $H_y(x,0)$ is
antisymmetric about the point $x = c = 0$.  Equations (\ref{tildeP}) and
(\ref{tildeQ}) then can be evaluated as
\begin{equation}
\tilde P(-\tilde b, \tilde b, \tilde \zeta) =
\Big(\frac{\tilde b^2  -\tilde \zeta^2 }{\tilde
w^2 -\tilde \zeta^2}\Big)^{1/2},
\label{tildePequal}
\end{equation} 
\begin{equation}
\tilde Q(-\tilde b, \tilde b, \tilde \zeta) = 2 \frac{\tilde \zeta}{\tilde
w}\Big[{\mathbf K}\Big(\frac{\tilde b}{\tilde w}\Big)+\frac{(\tilde w^2 - \tilde
\zeta^2)}{\tilde \zeta^2} {\mathbf \Pi}\Big(\frac{\tilde b^2}{\tilde
\zeta^2},\frac{\tilde b}{\tilde w}\Big)\Big],
\label{tildeQequal} 
\end{equation} 
where $\mathbf K(k)$ is the complete elliptic integral of the first kind of
modulus $k$.  In  (\ref{Hadomes}) and (\ref{Izdomes}),
$\tilde P(-\tilde b,
\tilde b, iL/\pi)$ is  pure real and  $\tilde Q(-\tilde b, \tilde b, iL/\pi)$
is pure imaginary, such that from  (\ref{hregionIV}) we obtain $\tilde h = 0$
when
$h= 0$.  Equations (\ref{K+w}) and (\ref{K-w}), resulting from the requirement that
$K_z(w)  = K_z(-w) = K_s$ at the critical current, reduce to a single equation,
\begin{equation}
 p \mathbf K(\tilde b') \sqrt{1-\tilde b'^2} = 1,
\label{tbph0} 
\end{equation} 
and the expression corresponding to  (\ref{icregionIV}) for the reduced
critical current $i_c$ becomes, with the help of  (\ref{tbph0}),
\begin{equation}
i_c = \sqrt{\frac{(1-\tilde b'^2)}{(1+\tilde b'^2 \tan^2 \theta)}}\mathbf
\Pi\Big(\frac{\tilde b'^2}{\cos^2 \theta + \tilde b'^2 \sin^2 \theta},
\tilde b'\Big) /\mathbf K(\tilde b')],
\label{icIV0}
\end{equation}
where $\theta = \pi w/L$.  
When $\tilde b' = 0$, we find that $p = 2/\pi$, and $i_c = 1$ for any value of
$\theta$. When
$\theta = 0$,  (\ref{icIV0}) reduces to 
\begin{equation}
i_c = \frac{\mathbf E(\tilde b')}{\mathbf K(\tilde b')\sqrt{1-\tilde
b'^2}}.
\label{icIV00}
\end{equation}
Plots of $H_y(x,0)$ and $K_z(x)$ vs $x$ for $h=0$ are similar to figure \ref{hyIV}
(a) and (b), except that $H_y(x,0)$ is an antisymmetric function
of
$x$, centered at the origin, and  $K_z(x)$ is a symmetric function of $x$.

\subsection{$h_1$, the boundary between regions III and IV}

For increasing values of the applied field $H_a$ (or the reduced field $h$), the
vortex dome expands and the antivortex dome shrinks.  For reduced fields $h$ in
the range $0 \le h < h_1$, we have $H_y(x,0) < 0$ for $x$ slightly larger than $a$. 
In other words, the coefficient of
$\tilde P_0(\tilde a,
\tilde b,
\tilde a + \epsilon)$ in  (\ref{Hy}) is negative.  
However, when $h = h_1$, this coefficient becomes zero.  For $h > h_1$, this
coefficient is positive, indicating that at the critical current, only a vortex
dome is present (region III in figure \ref{pvsh}).  The value of
$h_1$ is determined chiefly by the condition
that the  coefficient of
$\tilde P_0(\tilde a,
\tilde b,
\tilde a + \epsilon)$ in  (\ref{Hy}) is zero, which yields with the help of
 (\ref{tildeQb}), (17.7.7) of  \cite{Abram67}, and 
(414.01) of  
\cite{Byrd},
\begin{eqnarray}
&& \tilde h+ \frac{p \sec \theta}{(\tilde b'-\tilde a')\sqrt{(1-\tilde a')(1+\tilde
b')}}
\big[(1-\tilde a')(1+\tilde b') \mathbf E(\tilde q')
\nonumber \\
&&~~~~~~~~~~~~~~~-(1+\tilde a')(1-\tilde b') \mathbf K(\tilde
q') \nonumber \\ 
&&~~~~~~~~~-(\tilde b' - \tilde a')(1-\tilde b') \mathbf \Pi
(\frac{\tilde b' - \tilde a'}{1-\tilde a'},\tilde q')\Big]=0,
\label{h1}
\end{eqnarray}
where 
\begin{equation}
\tilde q'^2=\frac{2(\tilde b' - \tilde a')}{(1-\tilde a')(1+\tilde b')}.
\end{equation}
The value of $h_1$ for given values of $p > 2/\pi$ and $\theta = \pi w/L$ is
obtained as the value of $h$ that satisfies  (\ref{h1}), (\ref{K+w}), and
(\ref{K-w}).  The solutions of these equations also yield the values of $\tilde a'$
and $\tilde b'$ at the critical current when $h = h_1$.

\subsection{Region III: Two vortex-free zones and one vortex dome}

In region III of the diagram of $p$ vs $h$ shown in figure \ref{pvsh}, at the
critical current, the field and current distributions within each strip
divide into three zones.  For $b < x < w$, we have $H_y(x,0) = 0$ and
$K_z(x) > K_c$; for $ a < x < b$, we have  $H_y(x,0) > 0$ and $K_z(x) = K_c$; and
for $-w < x < a$, we have  we have $H_y(x,0) = 0$ and
$K_z(x) > K_c$.  Just above the critical current $I_c$, the voltage along the
length of the strip is produced by (a) small numbers of vortices nucleated at $x =
w$ as they overcome the geometrical edge barrier, followed by their  rapid
motion from
$x = w$ to $x = b$, (b) large numbers of vortices moving slowly 
through the vortex dome from
$x = b$ to $x = a$, and (c) small numbers of vortices leaking out of the vortex
dome and moving rapidly from $x = a$ to $x = -w$.  

\begin{figure}
\begin{center}
\includegraphics[width=8cm]{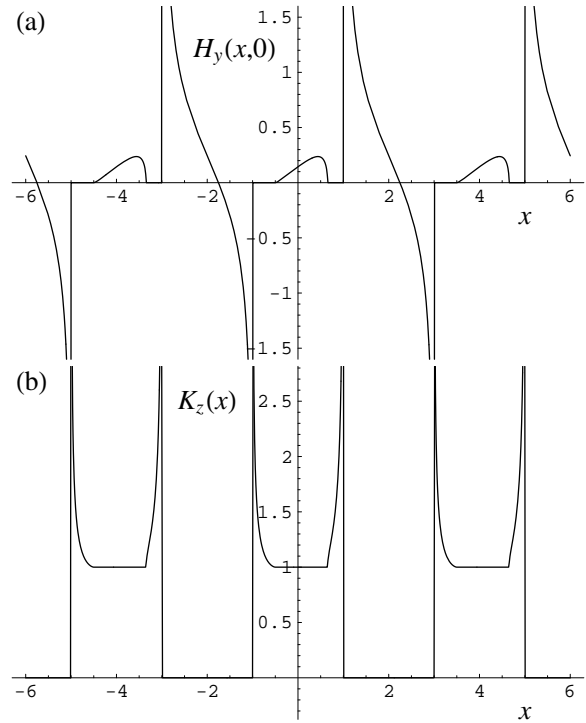}
\end{center}
\caption{Plots of (a) $H_y(x,0)$ (in units of $I_{s0}/2L\tan \theta$) and (b) 
$K_z(x)$ (in units of $K_p$) vs $x$ (in units
of
$w$) for $h = 0.2$, $p = 0.6$, and $\theta = \pi w /L = 0.5 \pi/2$ (region III),
for which $i_c = I_c/I_{s0} = 0.87$ and $I_c/I_p = 1.31$,
showing three strips ($-5 < x/w < -3,$
$-1 < x/w < 1,$ and $3 < x/w < 5$) and the gaps around them ($-7 < x/w < -5$, -$3
< x/w < -1$, $1 < x/w < 3,$ and $ 5 < x/w < 7)$.}
\label{hyIII}
\end{figure}

The four equations determining the values of $a$, $b$, $\tilde h$, and the reduced
critical current $i_c$  for given values of $h$, $p$ and $\theta$ in region III
are  (\ref{K+w}), (\ref{hregionIV}), (\ref{icregionIV}), and (\ref{h1}),
since the latter equation also can be shown to give the condition that
$dK_z(x)/dx = 0$ at $x = a$. Shown in figure \ref{hyIII}  (a) and (b)  are sample
plots of $H_y(x,0)$ and
$K_z(x)$ for $h$ and $p$ in region III.  
%***** Fig.5 ************************

\subsection{$h_2$, the boundary between regions III and II}

For increasing values of $h$, the left boundary of the vortex dome moves ever
closer to the left edge of the strip, i.e., $\tilde a \to \tilde w$ or $a \to -w$,
and numerical difficulties arise.  Although other criteria could be used, in
this paper we state that for practical purposes the
vortex-free zone on the left side shrinks to negligible width when
$\tilde a = -0.9999 \tilde w$.  For
given values of $p$ and $\theta$, this occurs at a reduced field $h = h_2$, which
can be obtained by solving  (\ref{K+w}), (\ref{h1}), and (\ref{hregionIV})
for $h$, $\tilde h$, and $b$ after replacing
$\tilde a$ by $ -0.9999 \tilde w$.
 
\subsection{Region II: One vortex-free zone and one vortex dome}

For values of $h > h_2$ in figure \ref{pvsh}, all the quantities
at the critical current can in principal be calculated using the same equations as
for Region III.  However, these quantities can be calculated with fewer numerical
difficulties and  with high accuracy by setting
$a = -w$ in the above equations, which results in a number of simplifications.  
The complex field still obeys $H(\zeta) = \tilde H(\tilde \zeta)$, but now, to
good approximation,
\begin{equation} 
\tilde H(\tilde \zeta) =\tilde P(-\tilde w, \tilde b, \tilde \zeta)
[\tilde H_a +
\frac{K_p}{2\pi}\tilde Q(-\tilde w,\tilde b,\tilde \zeta)],
\label{HzetaII}
\end{equation} 
where
\begin{equation}
\tilde P(-\tilde w, \tilde b, \tilde \zeta) =
\Big(\frac{\tilde \zeta-\tilde b}{\tilde \zeta-\tilde
w}\Big)^{1/2}
\label{tildePII}
\end{equation}
and
\begin{eqnarray} 
\tilde Q(\!-\tilde w,\tilde b,\tilde \zeta)\! &=&\!\!
\int_{-\tilde w}^{\tilde
b}\frac{\sqrt{\tilde w-\tilde u}}{(\tilde
\zeta-\tilde u)\sqrt{\tilde b-\tilde u}}d\tilde u 
\label{tildeQII} \\ &=&
\!2\sinh^{-1}\!\Big(\frac{\tilde w + \tilde b}{\tilde w - \tilde b}\Big)^{1/2}
\!\!\!\!-2\Big( \frac{\tilde \zeta\! - \tilde w}{\tilde \zeta\! - \!\tilde
b}\Big)^{1/2}\!\!\!\!
\sinh^{-1}\!\!\Big[\frac{(\tilde w\! +\! \tilde b)(\tilde \zeta \!-\! \tilde w)}
{(\tilde w\!-\!\tilde b)(\tilde \zeta \!+ \!\tilde w)}\Big]^{1/2}.
 \label{tildeQIIa}
\end{eqnarray}
This equation is equivalent to  (14b) of  \cite{Elistratov02}, which
was misprinted without the prefactor before the second $\sinh^{-1}$ term.

$H_y(x,0) = {\rm Re}H(x)$ and $K_z(x)
= -2 {\rm Im} H(x+i\epsilon)$ in region II are given by
\begin{eqnarray} 
H_y(x,0) &= &{\rm Re} \tilde H(\tilde x) \\
&=& \tilde P_0(-\tilde w, \tilde b, \tilde x)[\tilde H_a 
+\frac{K_p}{2 \pi} {\rm Re} \tilde Q(-\tilde w, \tilde b, \tilde x)], \\
\label{HyII} &&\;\;-\tilde w < \tilde x < \tilde b  \;{\rm or}\; |\tilde x| > \tilde
w, \nonumber \\
&=& 0,
~~~~~~~~~~ \tilde b < \tilde x < \tilde w.
\end{eqnarray}
\begin{eqnarray} 
K_z(x) & = & \tilde K_z(\tilde x) = -2{\rm Im} \tilde H(\tilde x+i\epsilon) \\
&=& K_p, \;-\tilde w < \tilde x < \tilde b,
\label{KzmiddleII} \\
&=& +2 \tilde P_0(-\tilde w, \tilde b, \tilde x)[\tilde H_a 
+\frac{K_p}{2 \pi}\tilde Q(-\tilde w, \tilde b, \tilde x)], \;\; \tilde b < \tilde
x < \tilde w,
\label{KzrightII} 
\end{eqnarray}
where 
\begin{equation}
\tilde P_0(-\tilde w, \tilde b, \tilde x) 
= \sqrt{\frac{|\tilde x - \tilde b|}{|\tilde x - \tilde
w|}}.
\label{tildeP0II}
\end{equation}

\begin{figure}%***** Fig.6 ************************
\begin{center}
\includegraphics[width=8cm]{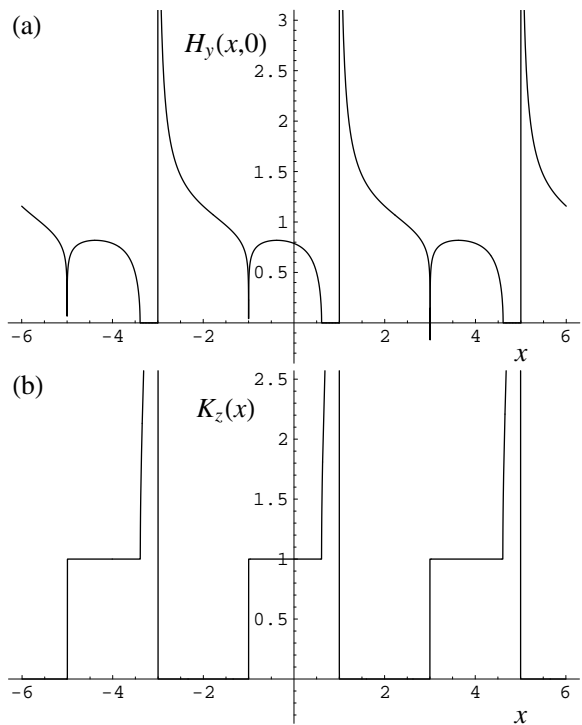}
\end{center}
\caption{Plots of (a) $H_y(x,0)$ (in units of $I_{s0}/2L\tan \theta$) and (b) 
$K_z(x)$ (in units of $K_p$) vs $x$ (in units
of
$w$) for $h = 1.0$, $p = 0.15$, and $\theta = \pi w /L = 0.5 \pi/2$ (region II),
for which $i_c = I_c/I_{s0} = 0.35$ and $I_c/I_p = 2.08$,
showing three strips ($-5 < x/w < -3,$
$-1 < x/w < 1,$ and $3 < x/w < 5$) and the gaps around them ($-7 < x/w < -5$, -$3
< x/w < -1$, $1 < x/w < 3,$ and $ 5 < x/w < 7)$.}
\label{hyII}
\end{figure}

Three equations must be solved simultaneously for $\tilde b$ (or $b$), $\tilde h$,
and
$i_c$ for given values of $h$, $p$, and $\theta$ in region II of figure
\ref{pvsh}.   The condition that  $K_z(w) = K_s$, evaluated as in deriving 
(\ref{K+w}), becomes
\begin{eqnarray}
&\sqrt{2(1-\tilde b')}\Bigg(\tilde h \cos \theta
+p\sinh^{-1}\sqrt{\frac{1+ \tilde b'}{1 - \tilde b'}}\Bigg) =1,
\label{K+wII}
\end{eqnarray}
and the other two equations needed are  (\ref{hregionIV}) and
(\ref{icregionIV}), in which 
the real quantities $ \tilde P_1,  \tilde P_2,  \tilde Q_1$ and
$
\tilde Q_2$ are defined via 
 $\tilde P(-\tilde w, \tilde b, iL/\pi)=\tilde P_1+ i
\tilde P_2$ and $\tilde Q(-\tilde w, \tilde b, iL/\pi)=\tilde Q_1+ i \tilde
Q_2$.
Shown in figure  \ref{hyII}  (a) and (b)  are sample
plots of $H_y(x,0)$ and
$K_z(x)$ for $h$ and $p$ in region II.  

\subsection{Critical current}

\begin{figure}%***** Fig.7 ************************
\begin{center}
\includegraphics[width=8cm]{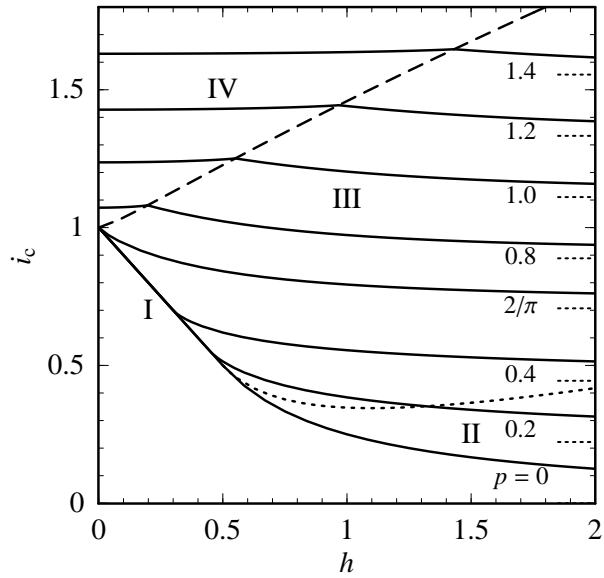}
\end{center}
\caption{$i_c(h,p,\theta)$ (critical current normalized to $I_{s0}$) vs reduced
field
$h$ for $\theta = 0.5 \pi/2$ and various values of the bulk pinning parameter
$p$. The solid curves exhibit 
$i_c$ vs $h$, the dashed curve shows $i_c$ at $h = h_1(p,\theta)$,
and  the dotted curve shows $i_c$ at $h = h_2(p,\theta)$.   For
$p<2/\pi$, $i_c$ decreases linearly with
$h$ (\ref{icI}) up to $h_d$ and then decreases more slowly in
regions III  and II.  One curve shows
$i_c$ for the special case of $p = 2/\pi$.  For $p>2/\pi$,
$i_c$ increases by a few percent in the double-dome region IV and then
decreases more gradually in regions III and II.  In all cases, $i_c$
asymptotically approaches $p \theta/\sin \theta$ for large $h$ (short dotted
lines along the right side of the figure).}
\label{icvsh}
\end{figure}

Using the above approach, one can calculate the normalized critical current $i_c =
I_c/I_{s0}$  (\ref{Is00}) for arbitrary values of the
reduced field
$h$  (\ref{h}), pinning parameter $p$ (\ref{p}), and $\theta = \pi
w/L$, where the strip width is $2w$ and the periodicity length is $L$.  Shown in
figure \ref{icvsh} are sample results for $\theta = 0.5 \pi/2$, when the widths of
the strips and the gaps are equal.  
Because of the effects of the geometrical
barrier, at sufficiently low applied fields and small values of $p$, the overall
critical current
$I_c$ can be considerably enhanced above the critical current $I_p = 2wK_p$ due to
bulk pinning alone.  This  occurs because the sheet-current density $K_z(x)$ in the
vortex-free zones near the sample edges can carry a supercurrent with a density
well above $K_p$.  However, in high applied fields the cross section of the
strip becomes nearly completely filled with vortices, and the sheet-current density
in the vortex-filled zones cannot exceed $K_p$.

\section{\label{conclusions}%
Conclusions}

In this paper we have considered an infinite array (X array) of coplanar equally
spaced superconducting strips and have calculated the critical current of each
strip in the presence of a perpendicular applied field, taking
into account not only geometrical barriers at the edges of the strips but also bulk
pinning characterized by a field-independent critical sheet-current density
$K_p$.  We have carried out these calculations using the X-array method of 
\cite{Mawatari96}, which enabled us to use solutions from 
\cite{Elistratov02} for an isolated strip.

Geometrical barriers at the edges of the strips enhance the critical current above
what it would be if it were due to bulk pinning alone.  At low fields, these
barriers delay the entrance of vortices and permit the strip edges to remain
vortex-free and  carry a sheet-current density well in excess of $K_p$.  Any
vortices and antivortices entering the strips arrange themselves into domelike
distributions, and under these domes the sheet-current density remains at
$K_p$.  As the perpendicular applied field increases, more and more vortices are
forced into the strip, causing the vortex domes to expand and the
high-current-density vortex-free zones to shrink, such that the overall critical
current  approaches that due to bulk pinning alone, $I_p = 2w K_p$.

The effects predicted in this paper for parallel arrays of narrow
superconducting strips would be most easily observed experimentally using
magneto-optical or scanning techniques in materials with low bulk pinning, such as
Bi-2212 (Bi$_2$Sr$_2$CaCu$_2$O$_{8+\delta}$), where geometrical-barrier effects were
first observed \cite{Schuster94,Zeldov94a}, or in low-pinning type-I
superconductors, such as Pb, where magnetic flux domes consisting of the
intermediate state have been observed \cite{Castro99}. However, it is important also
to consider the possibility of observing these effects in materials with strong
bulk pinning.

As a practical application of the above theory to coated conductors made of
superconducting YBCO (YBa$_2$Cu$_3$O$_{7-\delta}$), let us calculate the
critical-current enhancement due to geometrical barriers for an isolated strip and
then estimate how this enhancement is affected by striations.  We first consider
a long flat strip of  thickness $d$ = 1 $\mu$m and width $2w$ = 5 mm.  We assume
that its self-field bulk-pinning critical current density is $J_p$ = 1 MA/cm$^2$
(10$^{10}$ A/m$^2$), such that the bulk-pinnng critical sheet-current density is
$K_p = J_p d = 10^4$ A/m and the bulk-pinning critical current is $I_p = 2wK_p =
50$ A.  We use  (\ref{Is00}) with $K_s = 2H_s$, $\Lambda_c = d/2$, and $L \to
\infty$ to estimate the zero-field surface-barrier critical current $I_{s0}$.  We
use the experimental results of  \cite{Hao91} as discussed in 
\cite{Brojeny05}, from which we estimate that $H_{c1}$ = 180 Oe = 1.4
$\times 10^4$ A/m and $H_c$ = 3.2 kOe = 2.6 $\times 10^5$ A/m at 77 K.  We use the
conservative estimate that
$H_s = H_{c1}$, which yields $I_{s0}$ = $2 \pi H_{c1} \sqrt{dw}$  = 4.5 A.  The
corresponding scaling field
\begin {equation}
H_{scale} = I_{s0}/2L \tan \theta,
\label{Hscale}
\end{equation}
which appears in the denominator of  (\ref{h}) and (\ref{tildeh}), becomes
$H_{scale} = H_{c1} \sqrt{d/w} = $ 3.6 Oe = 286 A/m.  The corresponding value of
the pinning parameter is $ p = 11$, for which we obtain from  (\ref{tbph0})
and (\ref{icIV00}) that the  self-field  critical current $I_c$ at $h =
H_a/H_{scale} = 0$ is very nearly equal to $I_p$.  Moreover, $I_c$ remains very
nearly equal to to $I_p$ as the applied field increases. Our solutions show that
the vortex-free regions are very close to the edges at the critical current (i.e.,
$b$ and $-a$ are nearly equal to
$w$), such that the geometrical-barrier enhancement of
$I_c$ is negligible.

Next we suppose that the 5 mm strip is subdivided into 50 parallel strips, 
 each of width $2w$ = 98 $\mu$m, separated by gaps of
width 2 $\mu$m, with
period  $L$ = 100 $\mu$m.  To approximate the
behavior of the resulting striated conductor, we apply the above X-array results.  
Using 
(\ref{Is00}) with $K_s = 2H_s$, $\Lambda_c = d/2$, $\theta = \pi w/L = 0.49 \pi$,
and $\tan \theta = 31.8$, we find that the zero-field surface-barrier critical
current of one of the narrow strips is $I_{s0} = 2H_{c1}(\pi d L \tan
\theta)^{1/2}$ = 2.9 A.  The
corresponding bulk-pinning critical current is
$I_p = 2wK_p$ = 0.98 A, and the bulk-pinning parameter (\ref{p}) is $p$ =
0.22.  Since $p < 2/\pi$, the reduced critical current is $i_c = 1$, and the
self-field critical current of one of the strips is $I_c = I_{s0}$ = 2.9 A,
determined by the geometrical barrier alone.  Since there are 50 such strips, the
estimated total self-field critical current is 143 A.  The corresponding engineering
critical current density, taking the cross section to be 5 mm $\times$ 1 $\mu$m
(ignoring the cross section of the substrate) is $J_e = 2.9
\times 10^{10}$ A/m$^2$ = 2.9 MA/cm$^2$, as opposed to the 1 MA/am$^2$ critical
current density for the unstriated strip.  The reduced critical current $i_c$ vs
$h$ for this case is similar to that shown in figure \ref{icvsh} for $p$ = 0.2. 
However, the scaling field for
$i_c$
 vs $h$ from  (\ref{Hscale}) is  small: $H_{scale}$ = 5.7 Oe = 450 A/m.
To summarize, the X-array results using the above assumptions predict 
that the self-field critical current for the striated conductor can be
significantly enhanced 
above that due to bulk-pinning alone (by approximately a factor of three using the
above numbers).  However, the application of a relatively small perpendicular
magnetic field (of the order of tens of Oe or hundreds of A/m using the above
numbers) can be expected to produce a strong reduction of the enhancement and to
return the critical current nearly to that due to bulk-pinning alone.

Throughout this paper we have assumed that the bulk-pinning critical current density
$J_p$ is field-independent.  We now justify this assumption as follows.  We see from
figure \ref{icvsh} that the critical current has a significant dependence upon $h =
H_a/H_{scale} = B_a/B_{scale}$ only when $h \sim 1$.  In the above two paragraphs we
found for state-of-the-art YBCO coated conductors for which  $J_p \sim 1$ MA/cm$^2$
(10$^{10}$ A/m$^2$) that $H_{scale} < 10$ Oe = 796 A/m or $B_{scale} = \mu_0
H_{scale} <$ 10 G = 1 mT.  Experimentally it has been  found, for example in
\cite{Dam99, Gutierrez07}, that $J_p$ at 77 K in strong-pinning YBCO films is very
nearly independent of field for $H_a <$ 100 Oe $\approx$ 8000 A/m or $B_a < 100$ G =
10 mT.  Therefore it is an excellent approximation to assume that $J_p$ is
field-independent over the range of applied fields for which the effects of
geometrical barriers or edge pinning are relevant. 

In  \cite{Brojeny05}, we argued that in state-of-the art unstriated YBCO
coated conductors the pinning is due almost entirely to bulk pinning and that
geometrical barriers (or edge pinning) have a negligible effect upon the critical
current.  The above calculations indicate that for striated coated conductors the
additional edges produced by the striations could produce significant
enhancements of the critical current in self-field, but these enhancements are
strongly suppressed by relatively small applied magnetic fields.  In applications
where the magnetic flux density is of the order of 0.1 T or higher, where $J_p$
has a strong field dependence, we expect that geometrical barriers will have no
significant effect upon YBCO coated conductors, even if they are striated.

\ack{We thank  I.\ L.\ Maksimov for
stimulating discussions. This manuscript has been authored in
part by Iowa State University of Science  and Technology under Contract No.\
DE-AC02-07CH11358 with the U.S.\ Department of  Energy. }

\end{document}